\documentclass[aps,prb,twocolumn,superscriptaddress]{revtex4-1}
\usepackage{amssymb}
\usepackage{physics}
\usepackage{graphicx}
\usepackage{amsmath}
\usepackage{amsthm}
\usepackage{amsfonts}
\usepackage[T1]{fontenc}
\usepackage{array}
\usepackage{multirow}
\usepackage{color}
\usepackage{esint}
\usepackage{bm}
\usepackage{color}
\usepackage{bbm}
\usepackage{hyperref}
\usepackage{babel}
\usepackage{titlesec}
\usepackage{MnSymbol}

\newcommand{\Au}{ Aubry-Andr\'{e} }

\begin{document}
	\title{Generic mobility edges in several classes of duality-breaking one-dimensional quasiperiodic potentials}
	\author{DinhDuy Vu}
	\author{Sankar Das Sarma}
	\affiliation{Condensed Matter Theory Center and Joint Quantum Institute, Department of Physics, University of Maryland, College Park, Maryland 20742, USA}
	
	\begin{abstract}
		We obtain approximate solutions defining the mobility edge separating localized and extended states for several classes of generic one-dimensional quasiperiodic models. We validate our analytical ansatz with exact numerical calculations. Rather amazingly, we provide a single simple ansatz for the generic mobility edge, which is satisfied by quasiperiodic models involving many different types of nonsinusoidal incommensurate potentials as well as many different types of long-range hopping models. Our ansatz agrees precisely with the well-known limiting cases of the sinusoidal \Au model (which has no mobility edge) and the generalized \Au models (which have analytical mobility edges). Our work provides a practical tool for estimating the location of mobility edges in quasiperiodic systems. 
	\end{abstract}
	\maketitle
	
	\section{Introduction}
	Anderson localization \cite{Anderson1958, Abrahams1979, MacKinnon1981} is a great pillar of fundamental physics, where disorder may quantum mechanically suppress the coherent metallic transport leading to an insulating behavior as the extended states of the clean system become localized by the disorder. In three dimensions, Anderson localization allows for the mobility edge where the single-particle spectrum has a disorder-dependent critical energy separating the localized states from the extended states (and the localization leads to a metal-insulator transition as the Fermi level goes through the mobility edge). By contrast, disorder-induced Anderson localization is trivial in one dimension (1D) since any disorder localizes all states with no mobility edge and no metal-insulator transition.
	
	There are, however, situations where localization can emerge from a deterministic model. A prime example is when the Hamiltonian is incommensurate to the underlying 1D lattice, giving rise to the class of quasiperiodic systems. Such incommensuration, for example, can be incorporated into the onsite potential such that 
	\begin{equation}\label{AA}
		H=\sum_{i=1}^L\left(c_i^\dagger c_{i+1} + H.c.\right) + W\left(2\pi\beta i + \phi \right)c_i^\dagger c_i
	\end{equation}
	where the nearest-neighbor hopping energy and the lattice constant are chosen as units of energy and length, respectively. Here $W(x)=W(x+2\pi)$ is periodic but incommensurate with the underlying lattice ($i\in\mathbbm{Z}$) for an irrational $\beta$. The simplest such model is the \Au (AA) \cite{Aubry1980} with nearest-neighbor hopping and a quasiperiodic sinusoidal potential incommensurate with the lattice $W(2\pi \beta i + \phi) = V\cos(2\pi \beta i + \phi)$. The AA model is known to have a quasiperiodic potential-tuned localization transition at $V=2$ where all states are extended (localized) for $V<(>)2$. The quasiperiodic AA system is self-dual at the critical point $V=2$ with identical real space and momentum space representations, thus indicating an energy-independent localization transition for the whole spectrum with no mobility edges.
	
	It has been known since almost the beginning of the subject \cite{Soukoulis1982, Boers2007, Biddle2011, Bodyfelt2014, Liu2015, Gopalakrishnan2017, Rossignolo2019, Li2020, Wang2020, Liu2021, Wang2023} that breaking this fine-tuned AA duality (while still maintaining quasiperiodicity) leads to an energy-dependent self-duality relation, and hence, mobility edges. For example, adding a second incommensurate potential (e.g., a $\cos(4\pi\beta i)$ term) or adding a next-nearest-neighbor hopping term immediately takes the system away from the fine-tuned AA self-duality, producing a mobility edge. In particular, within a well-defined range of values of $V = \text{max}_x W(x)$, there could be a critical energy $E_c$, defining the mobility edge, separating extended states (localized) states for $E > (<) E_c$ for a given Hamiltonian. Such mobility edges in generic quasiperiodic problems are primarily obtained and studied numerically. However, there are a few well-known examples in the literature where analytical solutions have been derived for the mobility edge. One example is the model of Ref.~\cite{Biddle2010}, the Biddle-Das Sarma (BD) model, where the kinetic energy hopping term in the AA model (i.e., the first term in Eq.~\eqref{AA}) is long-ranged, but in an exponentially decaying spatial form. Another is the model of Ref.~\cite{Ganeshan2015}, the Ganeshan-Pixley-Das Sarma (GPD) model, where the quasiperiodic potential (i.e., the second term in Eq.~\eqref{AA}) is modified essentially to include all higher harmonics of the basic sinusoidal AA incommensurate potential, again leading to an energy-dependent analytically tractable self-duality producing mobility edges. In general, from a prior dual transformation, one can construct a model with analytic mobility edge \cite{Gopalakrishnan2017, Rossignolo2019, Li2020, Wang2020, Liu2021, Wang2023}. These models could be construed as the different generic possible classes representing generalized AA (GAA) models, where either the hopping kinetic energy or the incommensurate potential energy term is modified. We mention that the theoretically predicted mobility edges in GAA models have been experimentally observed \cite{Luschen2018, Kohlert2019, An2018, An2021}. However, a reversed problem, i.e., locating the mobility edge given a Hamiltonian is not well studied.
	
	In the current work, we present a rather unexpected theoretical result:  We provide an approximate simple-looking ansatz for two distinct generic classes of GAA models in one stroke, one with nearest-neighbor hopping as in Eq.~\eqref{AA}, but with non-sinusoidal potential comprising multiple powers of cosine potentials as the incommensurate term, and the other with the simple quasiperiodic potential of Eq.~\eqref{AA}, but with multiple long-range hopping terms. In the appropriate limits, our results reduce to the established analytical results of AA, BD, and GPD models, and we numerically validate our generic mobility edge results by comparing direct numerical simulations with our analytical theoretical ansatz. Although our theoretical ansatz is not exact, it tends to agree with the numerically calculated mobility edges up to a few percent deviations, even when the mobility edge has nontrivial and nonmonotonic structures in the parameter space of the quasiperiodic potential strength. Our finding can be directly experimentally verified and shows the richness of the quasiperiodic 1D localization compared with the simple random disorder-induced Anderson localization.
	
	Our work is deeply connected to a mathematical question of whether a self-adjoint operator has a pure point (bound state), absolutely continuous (free state), or singular continuous spectrum \cite{Avron1982, Simon1986, Del1994, Jitomirskaya1994, Gordon1997, Jitomirskaya1999,Jitomirskaya2012,Jitomirskaya2021}. Instead of focusing on the mathematical rigor, we aim at a more experimentally relevant goal, i.e., \textbf{an estimation} of mobility edge given a generic Hamiltonian. We mention that since our argument is based on duality, the ansatz should hold for most values of irrational $\beta$ and phase $\phi$, but there might exist some exceptions. This is similar to the AA model for $V>2$ that has a singular continuous spectrum (instead of pure point spectrum for localized states) if $\beta$ is a Liouville number or a dense set of $\phi$ if $W(x)=W(-x)$. We note that these exceptions form a zero-measure set mathematically, so the AA duality argument typically holds.
	
	\section{Hidden duality and Mobility edge ansatz}
	\begin{figure*}
		\centering
		\includegraphics[width=\textwidth]{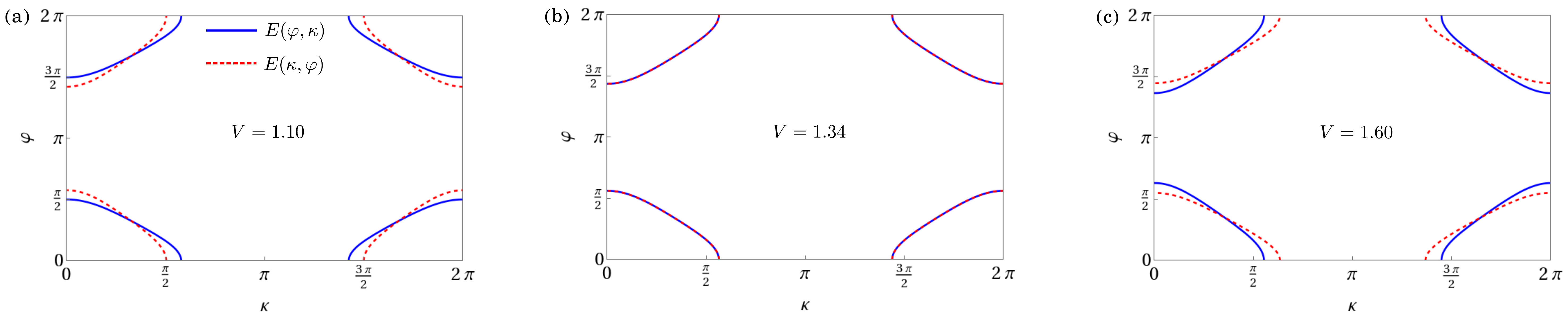}
		\caption{Fermi surfaces at $E=1.75,~\alpha_1=0.3,~\alpha_2=0.3$ with $\varphi,~\kappa$ and under the $\varphi\leftrightarrow\kappa$ exchange. At approximately $V=1.34$, the Fermi surface is invariant under the $\varphi\leftrightarrow\kappa$ exchange.}
		\label{fig1s}
	\end{figure*}	
	
	We consider a generic potential $W(2\pi\beta i + \phi)$ that is periodic under $2\pi$ translation $W(x)=W(x+2\pi)$. If $i\in\mathbbm{Z}$ (site index in tight-binding models) and $\beta$ is irrational, the potential never repeats itself exactly and the system is quasiperiodic. This makes directly solving the Hamiltonian difficult since one cannot use the ansatz of Bloch wavefunctions when $\beta$ is rational. One trick is to approximate $\beta\approx n_2/n_1$ (assuming $\beta<1$ so that $n_2<n_1$) and make the system size less than or equal to $n_2$. In that case, the potential is periodic under $n_1-$site translation so that one can use Bloch wave ansatz. However, the period is longer than the system size, which makes the system physically aperiodic. This method was used by Aubry and Andr\'e to derive the famous duality in the AA model, where self-dual points mark the localized-extended transition. To make any connection to the thermodynamic limit, one has to take the limit $n_2\to \infty$, which means the irrational $\beta$ is approximated with increasing accuracy. We note that $n_2$ does not appear explicitly in the AA duality, so the limit taking is only implied. 
	
	A similar duality, called hidden duality \cite{Gonccalves2022, Goncalves2023, Goncalves2023b}, can also be established in other quasiperiodic models using rational approximation. Assuming $\beta\approx n_2/n_1$, the lattice now has an enlarged unit cell of $n_1$ physical sites and can be solved in terms of rescaled Bloch momentum $\kappa=k/n_1 \in [-\pi,\pi]$. Because of the translational invariance under shifting $n_1$ sites, the phase $\phi$ of Eq.~\eqref{AA} is also rescaled into $\varphi=\phi/n_1\in [-\pi,\pi]$. We can then solve the Fermi surface in the 2D $(\kappa,\varphi)$ phase by solving the matrix
	\begin{widetext}
		\begin{equation}
			H(\kappa,\varphi) = \begin{pmatrix} 
				W(\varphi/n_1) & 1  & 0 & 0 & \dots & e^{i\kappa}\\
				1 & W(2\pi n_2/n_1 + \varphi/n_1) & 1 & 0 &\dots & 0\\   
				0 & 1 & W(2\pi n_2/n_1 + \varphi/n_1) & 1 &\dots & 0\\
				\dots\\
				e^{-i\kappa} & \dots & 0 & 0 & 1 & W[2\pi n_2(n_1-1)/n_1 + \varphi/n_1] 
			\end{pmatrix}
		\end{equation}
	\end{widetext}
	where $W(\phi)$ is the quasi-periodic potential except for the replacement $\beta$ with $n_2/n_1$.  
	
	It is hypothesized that in the limit $n_1, n_2\to\infty$ and at the self-dual point $(E,V)$ point, the Fermi surface is invariant under the exchange $(\cos\kappa \leftrightarrow \cos\varphi)$ or $(\cos\kappa \leftrightarrow -\cos\varphi)$. This duality is not guaranteed for finite $n_1$, but one can still numerically obtain a best-fit point that is most invariant under $\varphi\leftrightarrow\kappa$ exchange, which converges to the true self-dual point in the large $n_1$ limit \cite{Gonccalves2022, Goncalves2023, Goncalves2023b}. 
	
	The commensurate approximation can be tested analytically for AA and GPD models because the duality holds for any $n_1$. We start from the simplest $n_1 = 1$, then for AA potential $W(\varphi) = V\cos \varphi$
	\begin{equation}
		E(\kappa,\phi) = V\cos\varphi + 2\cos\kappa, 
	\end{equation}
	which is self-dual for $V=\pm 2$. For the GPD potential $W(\varphi) = V\cos\varphi/(1-\alpha\cos\varphi)$
	\begin{equation}
		\begin{split}
			&E(\kappa,\phi) = V\cos\varphi/\left(1-\alpha\cos\varphi\right) + 2\cos\kappa \\
			& \Leftrightarrow (E\alpha+V)\cos\varphi + 2\cos\kappa - 2\alpha\cos\varphi\cos\kappa - E = 0,
		\end{split}
	\end{equation}
	which is self-dual at $V = \pm 2 - E\alpha$.
	
	In the two examples above of analytical duality, the potential contains either a single cosine harmonic (AA model) or an exponential series (GDP model). In this paper, we are interested in a generic case given by a series of cosine with arbitrary coefficients
	\begin{equation}\label{genericV}
		W(\varphi) = V\cos\varphi\left(1+\sum_{l=1}^{m}\alpha_l \cos^l\varphi \right)	
	\end{equation}
	For the generic set of $\{\alpha_l\}$, the Fermi surface at $n_1-$rational approximant is a solution of
	\begin{equation}
		P_{n_1}[W(\varphi);E,V] + 2\cos\kappa = 0
	\end{equation}
	where $P_{n_1}[W(\varphi);E,V]$ is an $n_1-$order polynomial function of $W(\varphi)$ with $E$ and $V$ as parameters. For finite $n_1$, an exact dual point in general cannot be found because $\cos\kappa$ and $\cos\varphi$ exist in different powers; only for $n_1\to\infty$ that the polynomial function reduces to a different function, e.g., by Taylor series, that the exact duality is achieved. At $n_1=1$, we have
	\begin{equation}\label{eq:n1}
		\begin{split}
			E & = 2\cos\kappa  + V\cos\varphi\left(1+\sum_{l=1}^m \alpha_l \cos^l\varphi\right)
		\end{split}
	\end{equation}
	Instead of requiring the \textbf{entire Fermi surface} to be invariant under $\varphi\leftrightarrow\kappa$, which is impossible mathematically, we only impose this condition on \textbf{a pair of points} with either $\cos\varphi=0$ or $\cos\kappa=0$. For $\cos\varphi=0$, $\cos\kappa=E/2$, this point must be mapped to $\cos\varphi = E/2$ and $\cos\kappa = 0$, yielding
	\begin{equation}\label{condition}
		E = \frac{EV}{2}\left[1 + \sum_{l=1}^m \alpha_l \left(\frac{E}{2}\right)^l \right]
	\end{equation}
	and consequently, an approximated mobility edge
	\begin{equation}\label{ME}
		V = 2\left[1+\sum_{l=1}^m \alpha_l \left(\frac{E}{2}\right)^l\right]^{-1}.
	\end{equation}
	We note that the exchange $\cos\varphi\leftrightarrow - \cos\kappa$ yields another mobility edge for $V<0$, but for conciseness, we only focus on the $V>0$ side. The matching condition can be imposed on higher-order approximations of $\beta$ corresponding to larger $n_1$, which will most likely improve the mobility edge prediction. However, Eq.~\ref{ME} will become much more complicated with multiple powers of $\cos\kappa$ and $\cos\varphi$, and the corresponding theoretical prediction loses its simplicity and practical utility.\\
	
\begin{figure*}
		\centering
		\includegraphics[width=\textwidth]{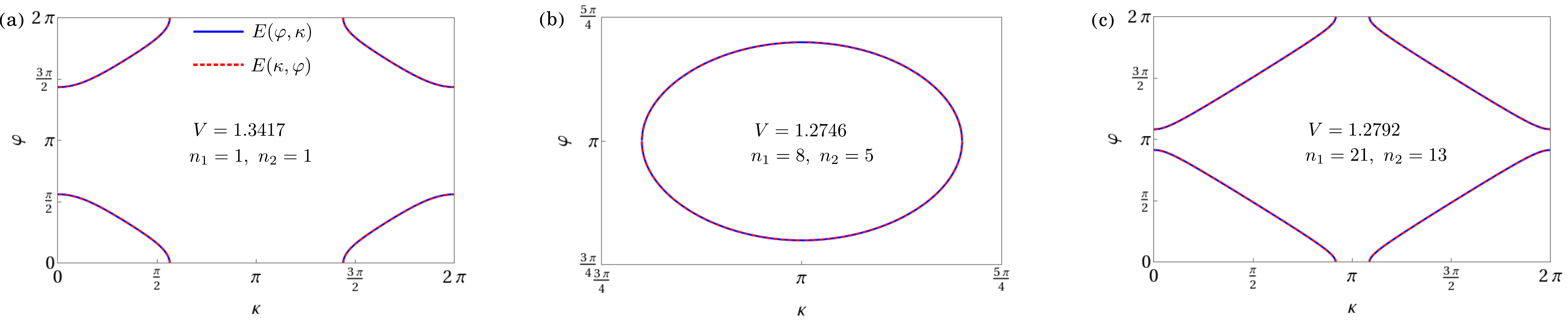}
		\caption{Fermi surfaces at $E=1.75,~\alpha_1=0.3,~\alpha_2=0.3$ with increasingly accurate approximation of the irrational period $\beta$. We expect in the limit of exact irrationality $n_1\to\infty$, the self-dual point locates at $V\approx 1.28$, which is 4\% away from our guess using Eq.~\eqref{ME}.}
		\label{fig2s}
	\end{figure*}

	We now benchmark our ansatz~\eqref{ME} against the numerical self-duality. For this purpose, we set $\beta=(\sqrt{5}-1)/2$ for simplicity so that $\beta$ can be progressively approximated by $F_m/F_{m+1}$ where $F_m$ is a Fibonacci number. We first check the ``numerical invariance''of the Fermi surface under $\varphi\leftrightarrow\kappa$ exchange at $n_1=1$ approximant. For $\alpha_1=\alpha_2=0.3$ and $E=1.75$, Eq.~\eqref{ME} predicts $V=1.34$. As shown in Fig.~\ref{fig1s}, this value produces a visibly invariant Fermi surface as compared to adjacent values. We then proceed with justifying the ansatz with increasing $n_1$. For each $n_1$, we fix $E$ and tune $V$ to optimize the $\varphi\leftrightarrow\kappa$-invariance. From Fig.~\ref{fig2s}, as $n_1\to\infty$, $V$ converges to a value that is only a few percent off our original guess. The reason our Fermi surface matching condition manifests so well numerically even though an exact $\varphi\leftrightarrow\kappa$ invariance is mathematically not possible for $n_1=1$ is the phase locking between different harmonics, i.e., the same phase $\phi$ is shared among all harmonics as shown in Eq.~\eqref{genericV}. When phase locking is relaxed, the Fermi surface loses its symmetric form and becomes more complicated. In that case, our simple ansatz based on $n_1=1$ will no longer well approximate self-dual points. 
	
	\begin{figure}
		\centering
		\includegraphics[width=0.5\textwidth]{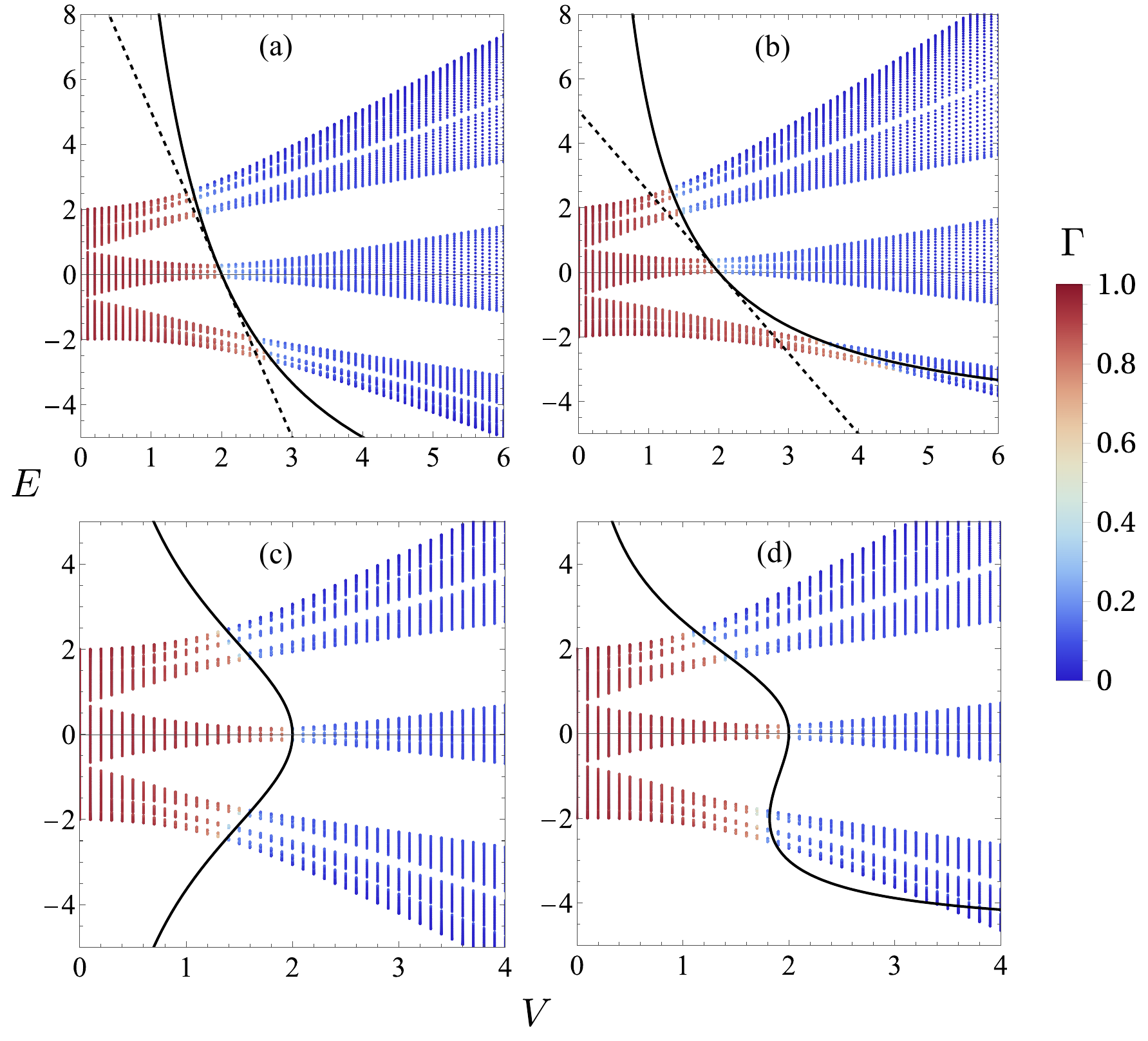}
		\caption{Fractal dimension spectrum and mobility edge predicted from Eq.~\eqref{ME}(solid lines) the GPD formula $V=2-\alpha_1E$ (dashed lines) (shown only in (a) and (b)) The coefficients of nonzero cosine harmonics are: (a) $\alpha_1=0.2$, (b) $\alpha_1=0.4$, (c) $\alpha_2=0.3$, (d) $\alpha_2=0.3, \alpha_3=0.2$. }
		\label{fig1}
	\end{figure}	
	
	\section{Numerical test on mobility edge ansatz}	
	
	We benchmark our approximate mobility edge against large-size numerical simulations using $\beta=(\sqrt{5}-1)/2$ and $L=2584$ in Eq.~\eqref{AA} with the periodic boundary condition. To quantify the localization degree of each eigenwavefunction and identify the mobility edge, we compute the fractal dimension \cite{Evers2008} defined by
	\begin{equation}
		\Gamma_j = \frac{-1}{\ln L} \ln \sum_{i=1}^L \bra{\psi_j}n_i\ket{\psi_j}^2
	\end{equation}
	with $\Gamma_j=0$ (1) for the maximally localized (extended) eigenstate $j$.	
	
	\subsection{Nearest-neighbor hopping and nonsinusoidal potential}
	
	We first justify our ansatz for a class of generalized AA models whose quasiperiodic models contain a series of higher powers of cosine given by Eq.~\eqref{genericV}. When all the $\alpha_l$ coefficients in this series are zero, we recover the simple sinusoidal quasiperiodicity of the AA model. For $\alpha_1$ nonzero, and all other $\alpha_l$ zero, we obtain the trichromatic incommensurate model numerically studied in Refs.~\cite{Soukoulis1982, Li2020}. In addition, Eq.~\eqref{genericV} reduces to the GPD model~\cite{Ganeshan2015} for exponentially decaying $\alpha_l=\alpha^l$, which has an exact mobility edge given by $V=2 - \alpha E$, i.e., $E_c = (2-V)/\alpha$.  We mention that the quasiperiodic potential in the GPD model is defined by $V \cos (2\pi \beta i + \phi) / [ 1 - \alpha \cos (2\pi \beta i + \phi)]$ \cite{Ganeshan2015}.	
	
	Before numerically testing the generic $\{\alpha_l\}$ case, we first discuss various limits of our ansatz, showing its agreement with the known cases of AA and GPD results.	
	
	The AA model follows from Eq.~\eqref{genericV} by putting all $\alpha_l=0$, which then gives the localization condition, according to our ansatz of Eq.~\eqref{ME}, to be $V=2$ (we focus on the positive solution) for all energies, i.e., no mobility edge-- all states are localized (extended) for $V > (<) 2$. This is, of course, the AA self-dual localization condition. The connection to the GPD model is that we set $\alpha_l=\alpha^l$, which then gives, according to our ansatz of Eq.~\eqref{ME}, the following localization condition:
	\begin{equation}
		V= 2\left[\sum_{l=0}^\infty \left(\frac{\alpha E}{2}\right)^l  \right]^{-1} = 2(1-\alpha E/2),
	\end{equation}
	leading to the mobility edge
	\begin{equation}
		E_c = (2-V)/\alpha.
	\end{equation}
	This is precisely the GPD model analytical mobility edge. Thus, our ansatz defined by Eq.~\eqref{ME} agrees with the limiting analytical results for AA and GPD models.
	
	To test how good our ansatz is for generic values of $\alpha_l$ for a completely general quasiperiodic potential, we show some numerical examples for a few representative situations with finite values of $\alpha_1$, $\alpha_2$, and $\alpha_3$ in Fig.~\ref{fig1}. It is obvious from Fig.~\ref{fig1} that our ansatz is surprisingly robust, providing $E_c$ as a function of $V$ with high accuracy. First, we only keep nonzero $\alpha_1$ so that the potential~\eqref{genericV} asymptotically approaches the GPD potential in the limit $\alpha_1\to 0$. This convergence reflects in Fig.~\ref{fig1}(a) for small $\alpha_1$; while for larger $\alpha_1$, as shown in Fig.~\ref{fig1}(b), our ansatz visibly outperforms the GPD formula. Remarkably, even the highly nontrivial reentrant localization structure with $E_c$ ($E_c$ being multivalued in Figs.~\ref{fig1}(c) and (d) for nonzero $\alpha_2$ and $\alpha_3$) is captured correctly by our simple ansatz! This shows that the applicable range of our ansatz is not simply a perturbative extension of known analytic solutions but, in fact, extends far beyond. We note that our ansatz is less accurate when $E$ approaches singularities of Eq.~\eqref{ME}. However, this regime tends to coincide with the edge of the spectrum [see Fig.~\ref{fig1}(b) and (d)] where the mobility edge, if exist here, will not have strong effect on measurements.
	
\begin{figure}
		\centering
		\includegraphics[width=0.5\textwidth]{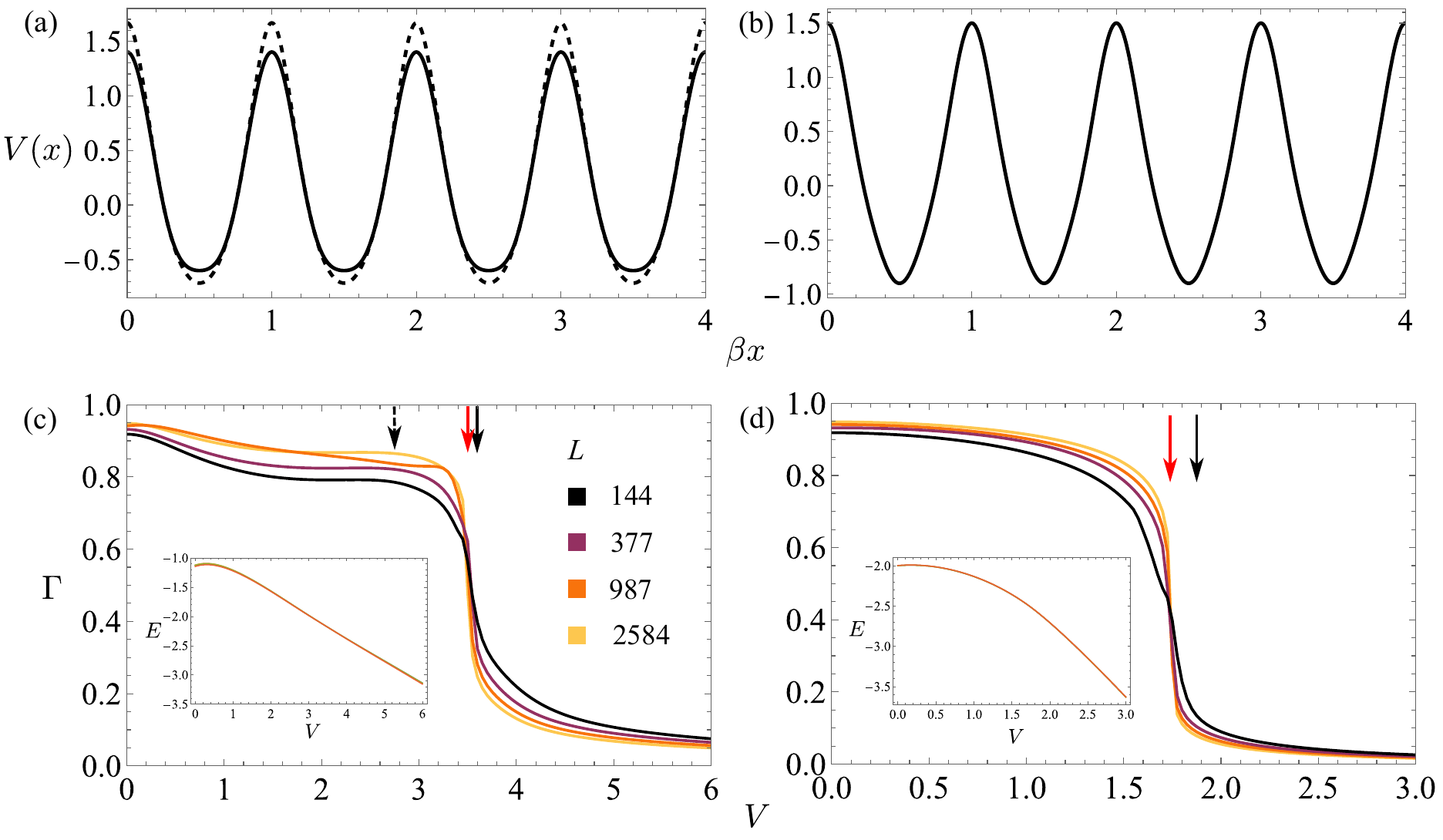}
		\caption{Top panel: incommensurate potential for $V=1$ (solid line) and GPD approximation if applicable (dashed line). Lower panel: Extended-localized phase transition along a path in the $(E, V)$ parametric phase shown in the inset obtained from finite-size scaling (red arrow), Eq.~\ref{ME} (black arrow), and approximate GPD if applicable (dashed arrow). (a), (c): $\alpha_1=0.4$. (b), (d): $\alpha_2=0.3,\alpha_3=0.2$. }
		\label{fig2}
	\end{figure}

	In Fig.~\ref{fig2}, we quantify the performance of our ansatz by comparing it with the critical point obtained from the standard finite-size scaling. Particularly, for $\alpha_1=0.4$ [Fig.~\ref{fig2}(a) and (c)], along a path in the $(E, V)$ parametric phase, Eq.~\ref{ME} incurs an error of $\sim 3\%$. At the same time, the next plausible theoretical prediction, the GPD model, suffers a much larger error of $\sim 20\%$. For a more complex non-sinusoidal incommensurate potential with $\alpha_2=0.3$ and $\alpha_3=0.2$ [Fig.~\ref{fig2}(b) and (d)], the error is only $\sim 7\%$. This is remarkable, given that this case has no other theoretical prediction or analytical solution.
	
	\subsection{Long-range hopping and sinusoidal potential}
	
	\begin{figure*}
		\centering
		\includegraphics[width=\textwidth]{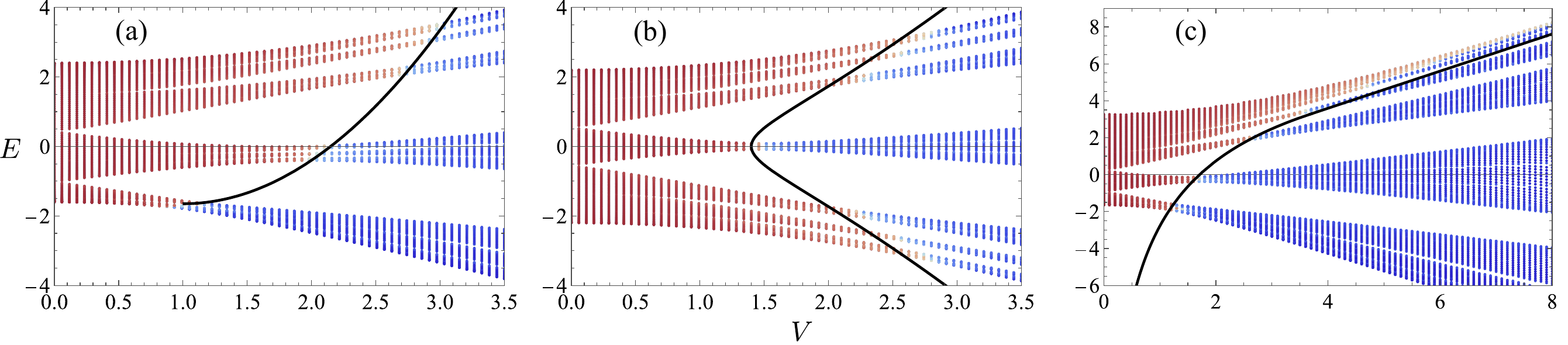}
		\caption{Fractal dimension spectrum with long-range hopping. (a) $t_2=0.2$, (b) $t_3=0.1$, (c) $t_l=1/l^2$.}
		\label{fig4}
	\end{figure*}	
	
	We now turn to the second class of generic models, which our theory captures correctly. These are models where the AA duality is broken by long-range (i.e., beyond nearest-neighbor) hopping terms with the incommensurate potential remaining the same as in the AA case.
	\begin{equation}
		\begin{split}
			H = \sum_i{\sum_{l=1}^m \left( t_l c_i^\dagger c_{i+l} + H.c. \right) + V\cos(2\pi\beta i + \phi)},
		\end{split}
	\end{equation}
	where we fix $t_1=1$ for consistence with other parts of the paper.
	For models with long-range hopping and a simple quasiperiodic sinusoidal potential, a Fourier transformation brings this long-range hopping model to a nearest-neighbor hopping model with a multi-harmonic nonsinusoidal incommensurate potential. At the $n_1=1$ approximation,
	\begin{equation}\label{longrangehopping}
		E = 2\sum_{l=1}^m t_l \cos (l\kappa)  + V\cos\varphi.
	\end{equation}
	
	In the BD model \cite{Biddle2010}, the long-range hopping case has an exact solution when the hopping strength decays exponentially with hopping distance, i.e., $t_l=e^{-\beta(l-1)}$ (which reduces to the AA model when $\beta\to\infty$, i.e., only the nearest-neighbor hopping $l=1$ has nonzero strength). Substituting this long-range hopping into Eq.~\eqref{longrangehopping}, we get the following
	\begin{equation}
		\begin{split}
			E =  e^\beta \left( -1 +\frac{\sinh \beta}{\cosh \beta -\cos \kappa} \right) + V\cos\varphi.
		\end{split}
	\end{equation}
	This immediately leads to
	\begin{equation}
		\cosh\beta\cos\varphi - e^\beta\cos\kappa\cos\varphi   + \frac{e^\beta+E}{V}\cos\kappa = \text{const} 
	\end{equation}
	and the corresponding linear mobility edge from $\cos\varphi\leftrightarrow \cos\kappa$ duality
	\begin{equation}
		V = \frac{E+e^\beta}{\cosh \beta}.
	\end{equation}
	This is precisely the mobility edge formula in BD~\cite{Biddle2010} for an exponentially decaying hopping amplitude with $1/\beta$ as the decay length.
	
	For a generic long-range hopping Hamiltonian, similar to the nonsinusoidal potential case, we cannot establish the exact duality at $n_1=1$. However, we can provide an approximate mobility edge by imposing the $\varphi\leftrightarrow\kappa$ invariance on a pair of points on the Fermi surface. The principle is completely identical to the nearest-neighbor hopping and nonsinusoidal potential case. Up to 3-site hopping, Eq.~\eqref{longrangehopping} becomes
	\begin{equation}\label{eq8}
		\begin{split}
			E &= V\cos\varphi + 2\cos\kappa + 2t_2\cos 2\kappa + 2t_3\cos 3\kappa \\
			&= V\cos\varphi + 2\cos\kappa + 2t_2(2\cos^2\kappa -1) \\
			&\quad + 2t_3(4\cos^3\kappa - 3\cos\kappa)
		\end{split}
	\end{equation}
	For $\cos\kappa=0$, $\cos\varphi = (E+2t_2)/V$. Imposing the duality between $(\cos\kappa, \cos\varphi) = (0,(E+2t_2)/V)$ and $((E+2t_2)/V,0)$, we obtain the equation
	\begin{equation}
		E=2\xi + 2t_2(2\xi^2-1) + 2t_3(4\xi^3-3\xi),~\xi = \frac{E+2t_2}{V},
	\end{equation}
	which is polynomial equation with multiple $V$ solutions for a fixed $E$. To filter unphysical solutions, we note that in the limit $t_n\to 0$, $V=2$, so we only use the real solution closest to 2. We compare this prediction with models having finite-range hoppings: $2-$site, and $3-$site hoppings in Fig.~\ref{fig4}(a) and (b), respectively, and observe a good agreement between the theoretical prediction and the numerical results. 
	
	We can also extend the mobility ansatz to a model with power-law decaying hopping $t_l=1/l^\gamma$ (we fix $t_1=1$) with $\gamma>1$ that is relevant to experiments on Rydberg atoms or spin qubits where the interaction (which can be mapped to fermionic hopping) is usually long-range. We note that the regime $1<\gamma<2$ can lead to unusual states that are both conducting and algebraically localized \cite{Saha2019}. Under the algebraically decaying hopping, Eq.~\eqref{longrangehopping} becomes
	\begin{equation}
		E = V\cos\varphi + \text{Li}_\gamma(e^{i\kappa}) + \text{Li}_\gamma(e^{-i\kappa}),
	\end{equation}
	where $\text{Li}_\gamma(z)$ is the polylogarithm. We can apply the same duality argument and draw an approximate mobility edge through the parametric equations
	\begin{equation}
		\begin{split}
			E(z) &= \text{Li}_\gamma(e^{i\arccos z}) + \text{Li}_\gamma(e^{-i\arccos z}),\\
			V(z) &= \frac{E(z)-E(0)}{z}.
		\end{split}
	\end{equation}
	We note that $E(z)$ is real only for $z\le 1$ which, in case of $\gamma=2$,
	set the constraint $E\le \pi^2/3\approx 3.2$. Our numerical result in the main text shows that the mobility edge extends beyond this limit. To continue the mobility edge, for $z>1$, we substitute the infinite sum by a finite sum up to the $100-$order term. This prediction also agrees with the numerical simulation shown in Fig.~\ref{fig4}(c).
	
	\subsection{Long-range hopping and nonsinusoidal potential}
	
	\begin{figure*}
		\centering
		\includegraphics[width=\textwidth]{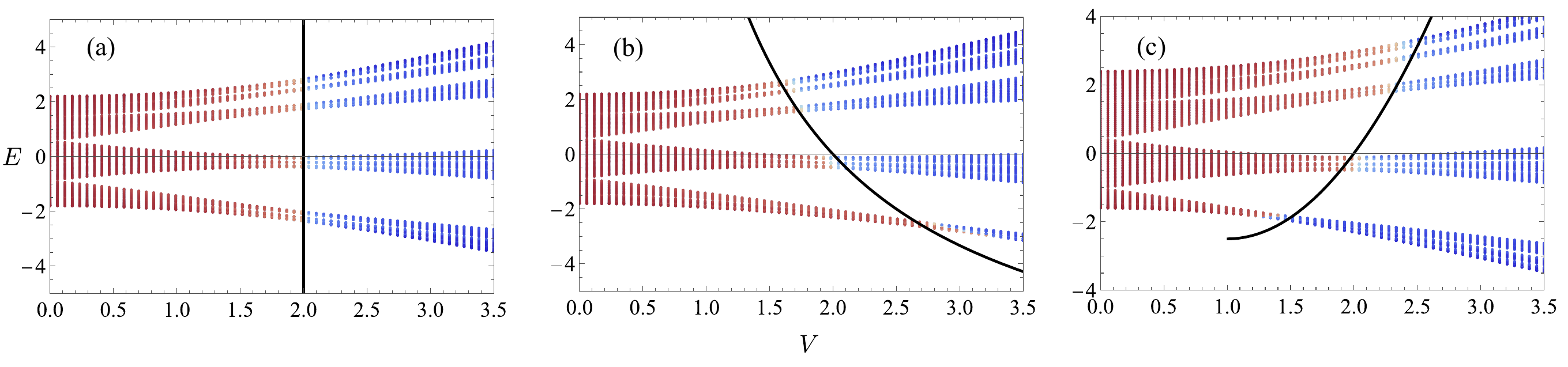}
		\caption{Fractal dimension spectrum with next-nearest hopping and second-harmonic potential, given the Hamiltonian~\eqref{combine}. (a) $t_2=\eta=0.1$, (b) $\eta=2t_2=0.2$, (c) $t_2=2\eta=0.2$.}
		\label{fig:combine}
	\end{figure*}	
	
	Finally, we consider a combined generic situation where both longer-range hopping and general nonsinusoidal quasiperiodicity are present with the Hamiltonian given by:
	\begin{equation}\label{combine}
		\begin{split}
			H&=\sum_i (c_i^\dagger c_{i+1} + t_2c_i^\dagger c_{i+2} + h.c.) \\
			&\quad + V\left[\cos(2\pi\beta i +\phi) + \eta \cos(4\pi\beta i + 2\phi)\right]c_i^\dagger c_i .
		\end{split}
	\end{equation}
	For simplicity, we include only two hopping terms and two quasiperiodic terms. Note that for $\eta=0$, the model of Eq.~\eqref{combine} reduces to the extensively numerically studied $t_1-t_2$ quasiperiodic model (with $t_1=1$ by choice here) \cite{Biddle2010, Li2020}.
	
	If $\eta=t_2$, we can use the Fourier transform to compute exactly the self-dual point, yielding $V=2$ for all energy. The situation is more complicated otherwise. The analog of Eqs.~\eqref{eq:n1} and \eqref{longrangehopping} is
	\begin{equation}
		E = V(\cos\varphi + \eta\cos 2\varphi) + 2(\cos\kappa + t_2\cos 2\kappa).
	\end{equation}
	Now, by setting either $\cos\varphi=0$ or $\cos\kappa=0$, we obtain two solutions, the matching of whom always yields the trivial condition: $\eta=t_2$ and $V=2$. To circumvent this situation, we note that in the limit $t_2\to 0$ ($\eta\to0$), only one solution for $\cos \varphi=0$ ($\cos\kappa=0$) is physical. We then enforce the duality in two cases.
	\begin{itemize}
		\item For $\eta > t_2$ which can be continuously deformed to the limit $\eta>0$ and $t_2 = 0$, at $\cos\varphi=0$, we only choose the solution
		\begin{equation}
			\cos \kappa = \frac{E+\eta V}{2} - t_2\left[2\left(\frac{E+\eta V}{2}\right)^2-1\right].
		\end{equation}
		This is also the solution of $\cos\varphi$ for $\cos\kappa=0$, which expresses the mobility edge. To further simplify the expression, we additionally assume that $V\approx 2$ and $1\gg \eta, t_2$, obtaining
		\begin{equation}\label{eq16}
			V = \frac{2}{1+(\eta-t_2)E}.
		\end{equation}
		
		\item For  $\eta < t_2$, we can simply obtain the solution by replacing $V\to 2/V$, $E\to 2E/V$ and $\eta \leftrightarrow t_2$ in Eq.~\eqref{eq16}
		\begin{equation}
			\begin{split}
				\frac{2}{V} &= \frac{2}{1+2(t_2-\eta)E/V} \\
				\Rightarrow V &= 1+\sqrt{1+4(t_2-\eta)E}.
			\end{split}
		\end{equation}
	\end{itemize}
	In either case, the solution reduces to the AA self-dual critical point $V=2$ for $t_2=\eta$, consistent with the argument above based on the Fourier transformation. We show the corresponding numerical results in Fig.~\ref{fig:combine}.

	\section{Conclusion}
	We introduce in this work a simple ansatz, extensively validated by exact numerical diagonalization, for the mobility edges in several generic classes of 1D quasiperiodic localization models where the generic quasiperiodicity is nonsinusoidal, and the hopping is long-ranged. Our ansatz matches several existing fine-tuned limiting cases with analytic mobility edge. However, more importantly, our ansatz agrees exceptionally well with exact numerical results throughout the entire parameter space. This rather surprising finding of an approximate generic analytical solution for several classes of generic quasiperiodic models hints at the complex richness of quasiperiodic localization. Our results are easily verifiable in experiments using atomic gases and optical lattices where many fine-tuned quasiperiodic localization models have already been studied \cite{Roati2008, Schreiber2015, Luschen2018, Kohlert2019, An2018, An2021}. Quasiperiodic modulations also manifest naturally in Moir\'{e} systems \cite{Huang2019, Carr2020, Gonccalves2021, Vu2021} where our work is also relevant.
	
	We note that experimental systems are always finite-size, which has significant implications on the role of the irrational frequency $\beta$. In the Appendix, we show examples where $\beta = L/Q$ where $Q$ and $L$ are coprime numbers and $L$ is the system size. Our ansatz does not hold if $Q$ is too close to $L$, i.e., the system is mathematically aperiodic, but the electron has to travel a long distance to register this property. This is also the intuition behind the breakdown of localization in the AA model for $V>2$ and almost-rational Liouville-number frequency.

	\begin{acknowledgments}
		\textit{Acknowledgments - }
		This work is supported by Laboratory for Physical Sciences.
	\end{acknowledgments}
	
	\begin{figure}
	\centering
	\includegraphics[width=0.5\textwidth]{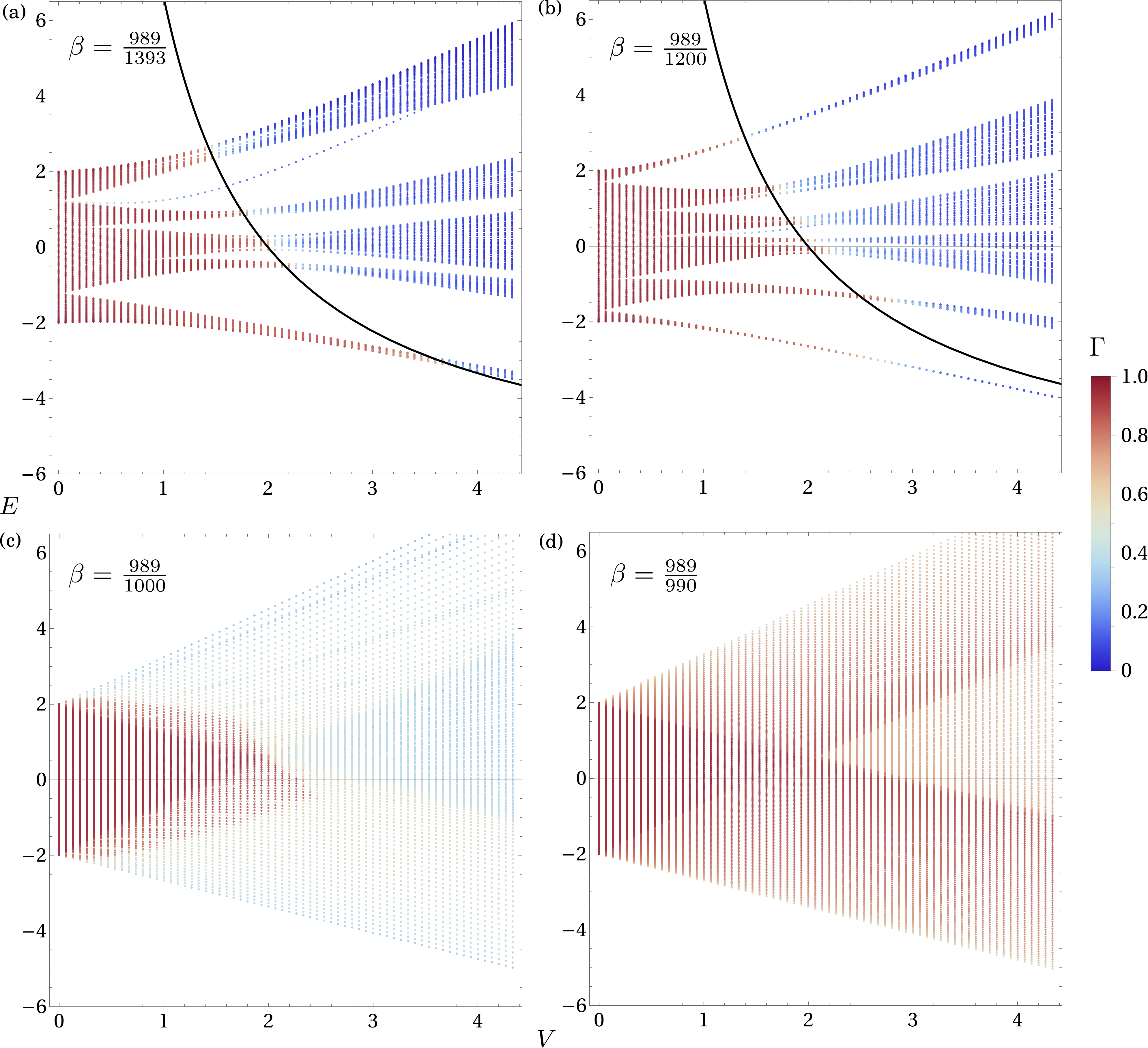}
	\caption{Fractal dimension spectrum in the nearest-neighbor hopping and nonsinusoidal model with $\alpha_1=0.3$. The system size is 989, and the frequency $\beta$ is a rational number whose denominator is larger than the system size. Solid lines are drawn from Eq.~\eqref{ME} when applicable.}
	\label{fig:appendix}
\end{figure}		
	
	\bibliographystyle{apsrev4-2}
	\bibliography{SPME}
	
	\appendix
	\section{Numerical results for almost-periodic finite-size systems}
	In this appendix, we provide extra data on the finite-size effect on our ansatz. In the main text, we fix the system size to be a Fibonacci number $F_m$ so that we can simulate an aperiodic system by choosing  $\beta = F_m/Q$ where $Q$ and $F_m$ are coprime numbers and $Q>F_m$. If we choose $Q$ to be the following Fibonacci number $F_{m+1}$, $\beta$ approximates the golden ratio that we use in the main text, and our ansatz agrees well with the numerical result shown in Fig.~\ref{fig:appendix}(a). For $Q$ close to $F_m$ [Fig.~\ref{fig:appendix}(c-d)], the supposedly localized regime becomes delocalized with the scaling exponent $\Gamma$ being either fractional ($0<\Gamma<1$) or close to unity. Therefore, our ansatz breaks down if the system's finite size is insufficient to resolve the quasiperiodicity. 
	
\end{document}